\documentclass[prb,twocolumn,showpacs,amsmath,superscriptaddress,psfig,amssymb]{revtex4}

\usepackage{graphicx}
\usepackage{dcolumn}
\usepackage{bm}

\begin{document}


\title{The upper critical field and its anisotropy in LiFeAs}

\author{J. L. Zhang}
\affiliation{Department of Physics, Zhejiang University, Hangzhou,
Zhejiang 310027, China}
\author{L. Jiao}
\affiliation{Department of Physics, Zhejiang University, Hangzhou,
Zhejiang 310027, China}
\author{F. F. Balakirev}
\affiliation{NHMFL, Los Alamos National Laboratory, MS E536, Los
Alamos, NM 87545, USA}
\author{X. C. Wang}
\affiliation{Beijing National Laboratory for Condensed Matter
Physics,Institute of Physics, Chinese Academy of Science, Beijing,
100080, China}
\author{C. Q. Jin}
\affiliation{Beijing National Laboratory for Condensed Matter
Physics,Institute of Physics, Chinese Academy of Science, Beijing,
100080, China}
\author{H. Q. Yuan}
\email{hqyuan@zju.edu.cn} \affiliation{Department of Physics,
Zhejiang University, Hangzhou, Zhejiang 310027, China}

\date{\today}

\begin{abstract}

The upper critical field $\mu_0H_{c2}(T_c)$ of LiFeAs single
crystals has been determined by measuring the electrical resistivity
using the facilities of pulsed magnetic field at Los Alamos. We
found that $\mu_0H_{c2}(T_c)$ of LiFeAs shows a moderate anisotropy
among the layered iron-based superconductors; its anisotropic
parameter $\gamma$ monotonically decreases with decreasing
temperature and approaches $\gamma\simeq 1.5$ as $T\rightarrow 0$.
The upper critical field reaches 15T ($H\parallel c$) and 24.2T
($H\parallel ab$) at $T=$1.4K, which value is much smaller than
other iron-based high $T_c$ superconductors. The temperature
dependence of $\mu_0H_{c2}(T_c)$ can be described by the
Werthamer-Helfand-Hohenberg (WHH) method, showing orbitally and
(likely) spin-paramagnetically limited upper critical field for
$H\parallel c$ and $H\parallel ab$, respectively.

\end{abstract}


\maketitle

\section{Introduction}

The discovery of superconductivity in iron pnictides\cite{Kamihara}
has attracted world-wide interests in searching for new type of high
$T_c$ superconductors and unveiling their unconventional nature of
superconductivity. Until now, several series of iron-based
superconductors have been found \cite{zhao 2009,Lumsden}, which
posses a similar layered crystal structure to those of the high
$T_c$ cuprates. Resembling the cuprates and heavy fermions,
superconductivity in most of the iron pnictides/chalcogenides seems
to be closely tied up with magnetism  \cite{zhao 2009,Lumsden};
superconductivity appears while antiferromagnetism is suppressed by
hole (or electron) doping or by application of external pressure. In
particular, the layered crystal structure and the high
superconducting transition temperatures of the iron
pnictides/chalcogenides initially suggested a strong analogy with
the cuprates, providing an alternative to study the puzzles of high
$T_c$ superconductivity. However, significant discrepancies have
been observed between the iron-based superconductors and other
layered superconductors. For example, d-wave superconductivity was
realized in the high $T_c$ cuprates, but an $s\pm$-type order
parameter has been proposed for the iron pnictides/chalcogenides
superconductors \cite{Hu,Mazin,Hanaguri}. Upper critical field is
another important superconducting parameter. A large upper critical
field has been identified in both iron pnictides/chalcogenides and
the cuprates, but the former shows a rather weak effect of
anisotropy \cite{Yuan,Fang,Jaroszynski}. In particular, nearly
isotropic upper critical field $\mu_0H_{c2}(T_c)$ has been observed
in the 122- and 11-type iron pnictides/chalcogenides
\cite{Yuan,Fang}, remarkably different from any other layered
superconductors. LiFeAs, a much cleaner compound with a large ratio
of room temperature resistivity to residual resistivity (RRR$\sim$
40), seems to be very unique among the iron pnictide superconductors
\cite{Wang,Tapp,Yoo}. Bearing a nearly identical structure of
(Fe$_2$As$_2$)$^{2-}$ and also a similar electronic structure to
other iron pnictides \cite{Nekrasov}, LiFeAs, however, shows simple
metallic behavior prior to entering the superconducting state,
lacking evidence of structural/magnetic transitions. Moreover, the
stoichiometric compound LiFeAs becomes superconducting at ambient
pressure and without introducing additional charge carriers via
doping. Nevertheless, LiFeAs still demonstrates a relatively high
$T_c$ ($T_c\simeq 18$ K), being comparable with those iron
pnictides/chalcogenides which parent compounds undergo a
magnetic/structural transition. Unfortunately, LiFeAs is extremely
air sensitive and many of its superconducting properties remain
mysterious because of the restrictions of accessible experimental
methods. In LiFeAs the extrapolation of $\mu_0H_{c2}$ near $T_c$ to
zero temperature gives a rather large value of $\mu_0H_{c2}(0) (\sim
80$ T) \cite{Yoo}. In order to fully track the field dependence of
superconductivity, a strong magnetic field is desired. Here we
report the first resistivity measurement of LiFeAs in a pulsed
magnetic field down to 1.4K, from which the temperature-magnetic
field phase diagram is well established. The upper critical field
$\mu_0H_{c2}$ is determined to be 15 T and 24.2 T at $T=1.4$K for
$H\parallel c$ and $H\parallel ab$, respectively. In comparison with
other series of iron pnictide superconductors, the upper critical
field shows a moderate anisotropic effect and its value of
$\mu_0H_{c2}(0)$ is largely reduced.

\section{Experimental methods}

 High-quality single crystals of LiFeAs have been grown by a self-flux technique \cite{Wang}. The
precursor of Li$_3$As was synthesized from Li piece and As chips
that were sealed in a Nb tube under Ar atmosphere and then treated
at 650$^\circ$C for 15 hours in a sealed quartz tube. The Li$_3$As,
Fe and As powders were mixed in the ratio of Li:Fe:As=1:0.8:1. The
powder mixture was then pressed into a pallet in an alumina oxide
tube. To prevent the vaporized Li from attacking the quartz tube at
high temperature, the sample pallet was subsequently sealed in a Nb
tube and a quartz tube under vacuum. The sealed quartz tube was
heated at 800$^\circ$C for 10h before heating up to 1100 $^\circ$C
at which it was hold for another 10h. Finally, it was cooled down to
800$^\circ$C with a rate of 5$^\circ$C per hour. Crystals with a
size up to 4mm$\times$3mm$\times$0.5mm were obtained. The whole
preparation work were carried out in a glove box protected with high
purity Ar gas. The obtained single crystals were first characterized
by x-ray diffraction with a Mac Science diffractometer and ac
susceptibility measurements using the Oxford cryogenic system
(Maglab-Exa-12) prior to the transport measurements in a pulsed
magnetic field at Los Alamos.

\begin{figure}[b]\centering
 \includegraphics[width=8cm]{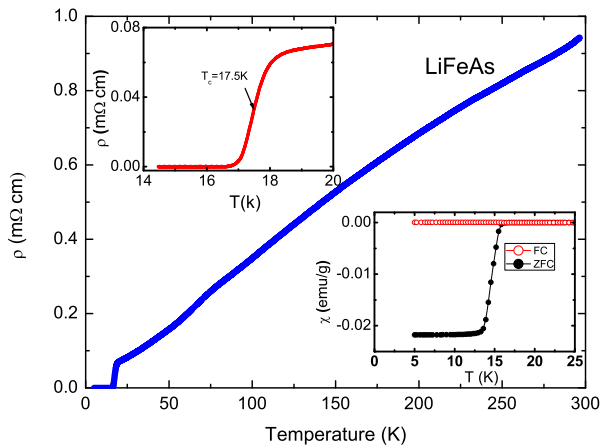}
\caption{Temperature dependence of the electrical resistivity
$\rho(T)$ for LiFeAs at zero field. The lower inset shows the
magnetic susceptibility $\chi(T)$.}\label{fig1}
 \end{figure}

Electrical resistivity was measured using a typical four-contact
method in pulsed fields of up to 40T and at temperatures down to
1.4K in a Helium-4 cryostat. Note that the applied electrical
current was always along the ab-plane. In order to minimize the
inductive self-heating caused by the fast change of magnetic field,
small crystals with typical sizes 2mm $\times$0.5mm$\times$0.1mm
were cleaved off along the c-direction from the as-grown samples. In
order to avoid oxidizing the samples, special cares were paid to
protect the samples from exposing to air while preparing for the
electrical contacts. Data were recorded using a 10~MHz digitizer and
100~kHz alternating current, and analyzed using a custom low-noise
digital lock-in technique. Temperature dependence of the electrical
resistivity at zero field was measured with a Lakeshore resistance
bridge.

\begin{figure}[b]\centering
 \includegraphics[width=8cm]{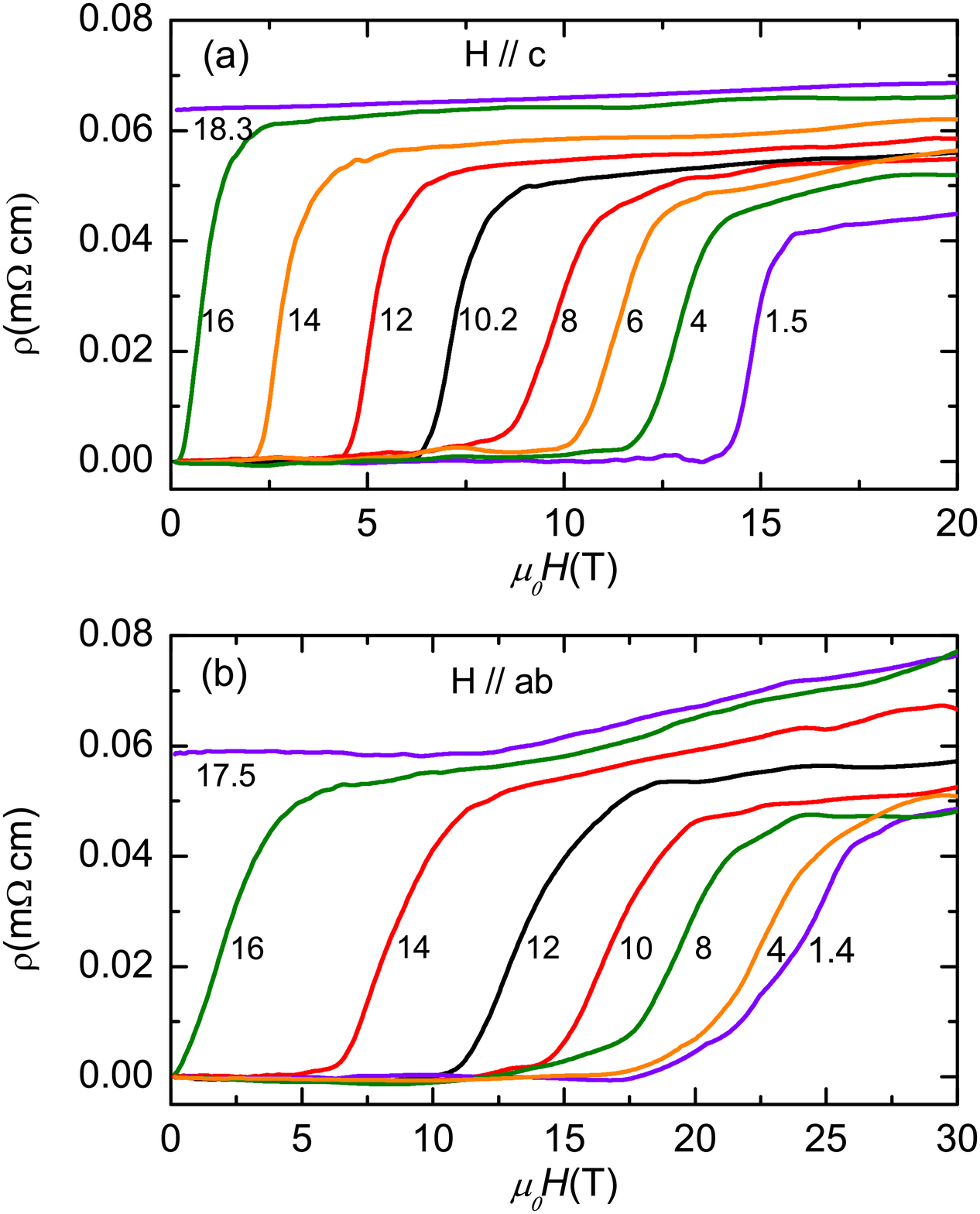}
\caption{Magnetic field dependence of the electrical resistivity at
variant temperatures for LiFeAs: (a) $H\parallel c$; (b) $H\parallel
ab$.}\label{fig2}
 \end{figure}

\section{Experimental results and discussion}

Fig. 1 presents the temperature dependence of the electrical
resistivity $\rho(T)$ at zero magnetic field for LiFeAs. Obviously,
LiFeAs shows simple metallic behavior upon cooling down from room
temperature, followed by a sharp superconducting transition at
$T_c\simeq 17.5$ K, which is in consistence with the reports in
literature \cite{Wang,Tapp,Yoo}. Note that the weak kink in the
resistivity $\rho(T)$ around 75K is attributed to the change of
cooling rate. No evidence of structural/magnetic transition has been
observed in LiFeAs. In order to demonstrate the superconducting
transition in detail, we plot the low temperature electrical
resistivity and magnetic susceptibility in the inset of Fig.1, which
were measured with samples cut from the same batch. As frequently
observed in superconductors, the bulk $T_c$ determined from the
magnetic susceptibility is slightly lower. The observations of a
large RRR ($\simeq15$) and a narrow superconducting transition
indicate high quality of the samples investigated here.
Since LiFeAs
is a good metal with low resistivity, measurements of its electrical
resistivity in a pulsed magnetic field is rather challenging.
Nevertheless, we have succeeded in obtaining a good set of
resistivity data up to a magnetic field of 40T after many failures.
Fig. 2 shows the field dependence of the electrical resistivity
$\rho(\mu_0H)$ of LiFeAs at variant temperatures, in which the
magnetic field is applied along (a) the c-axis and (b) the ab-plane,
respectively. One can see that a relatively sharp superconducting
transition survives down to very low temperatures, even though the
signals become more noisy upon cooling down, in particular for the
case of $H\parallel ab$. Obviously, the superconducting transition
is eventually suppressed upon applying a magnetic field, but the
critical field required to suppress superconductivity is much larger
for $H\parallel ab$. Furthermore, the normal state of LiFeAs remains
metallic upon suppressing superconductivity in a sufficiently high
magnetic field, being different from those of the 122- and 11-type
compounds\cite{Yuan,Fang}.

\begin{figure}[htbp]\centering
 \includegraphics[width=8cm]{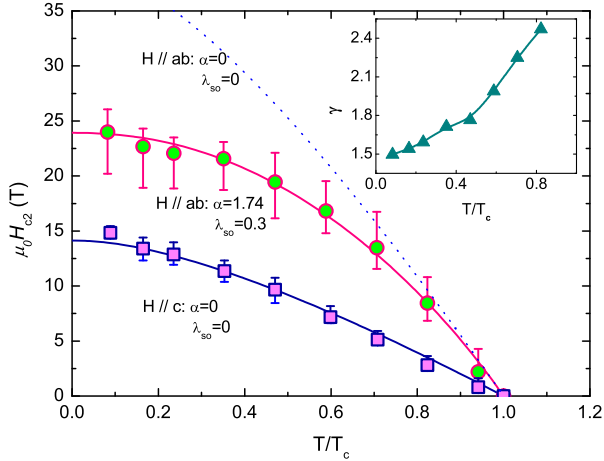}
\caption{The upper critical field $\mu_0H_{c2}(T_c)$ and the
corresponding WHH fits for LiFeAs. The solid lines are the best fits
to the experimental data and the dotted line is the WHH fit without
considering the spin paramagnetic effect. The inset shows the
temperature dependence of the anisotropic parameter
 $\gamma$.}\label{fig3}
 \end{figure}

The upper critical field $\mu_0H_{c2}(T_c)$ of LiFeAs, determined
from the mid-point of the superconducting transition, is plotted in
Fig. 3. The error bars were derived from the 20\% and 80\% drop of
the normal state resistivity at $T_c$. In comparison with other
families of the iron-based high temperature superconductors
\cite{Yuan,Fang,Jaroszynski}, LiFeAs shows a relatively small upper
critical field, reaching $\mu_0H_{c2}$=15T and 24.2T at $T=1.4$K for
$H\parallel c$ and $H\parallel ab$, respectively. Temperature
dependence of the anisotropic parameter, defined as
$\gamma=H_{c2}^{H\parallel ab}/H_{c2}^{H\parallel c}$, is plotted in
the inset of Fig.3. Resembling those of the previously investigated
iron-based superconductors\cite{Yuan,Fang,Jaroszynski}, the
anisotropic parameter $\gamma$ decreases with decreasing
temperature, reaching $\gamma=1.5$ at zero temperature. Such a value
of $\gamma$ is slightly higher than that of the 122- and 11-type
compound \cite{Yuan,Fang}, which shows nearly isotropic behavior at
low temperatures, but significantly smaller compared to that of the
high $T_c$ cuprates and organic superconductors \cite{shriffer,
singleton}.
According to the Werthamer-Helfand-Hohenberg method
\cite{WHH}, the upper critical field limited by the orbital
mechanisms in the dirty limit is given by:

\begin{equation}
\mu_0
H_{c2}^{orb}(0)[\rm{T}]=-0.69\it{T_c(dH_{c2}/dT)\mid_{T=T_c}}[\rm
K]. \label{eq:one}
\end{equation}

On the other hand, superconductivity is suppressed while the
magnetic energy associated with the Pauli spin susceptibility in the
normal state exceeds the condensation energy in the superconducting
state as a result of Zeeman effect. In this case, the Pauli-limited
upper critical field for weakly coupled superconductors can be
written as \cite{Clogston,Chandrasekhar}:

\begin{equation}
\mu_0 H_{c2}^P(0)[\rm{T}]=1.86\it{T_c}[\rm K]. \label{eq:two}
\end{equation}

\begin{figure}[b]\centering
 \includegraphics[width=8cm]{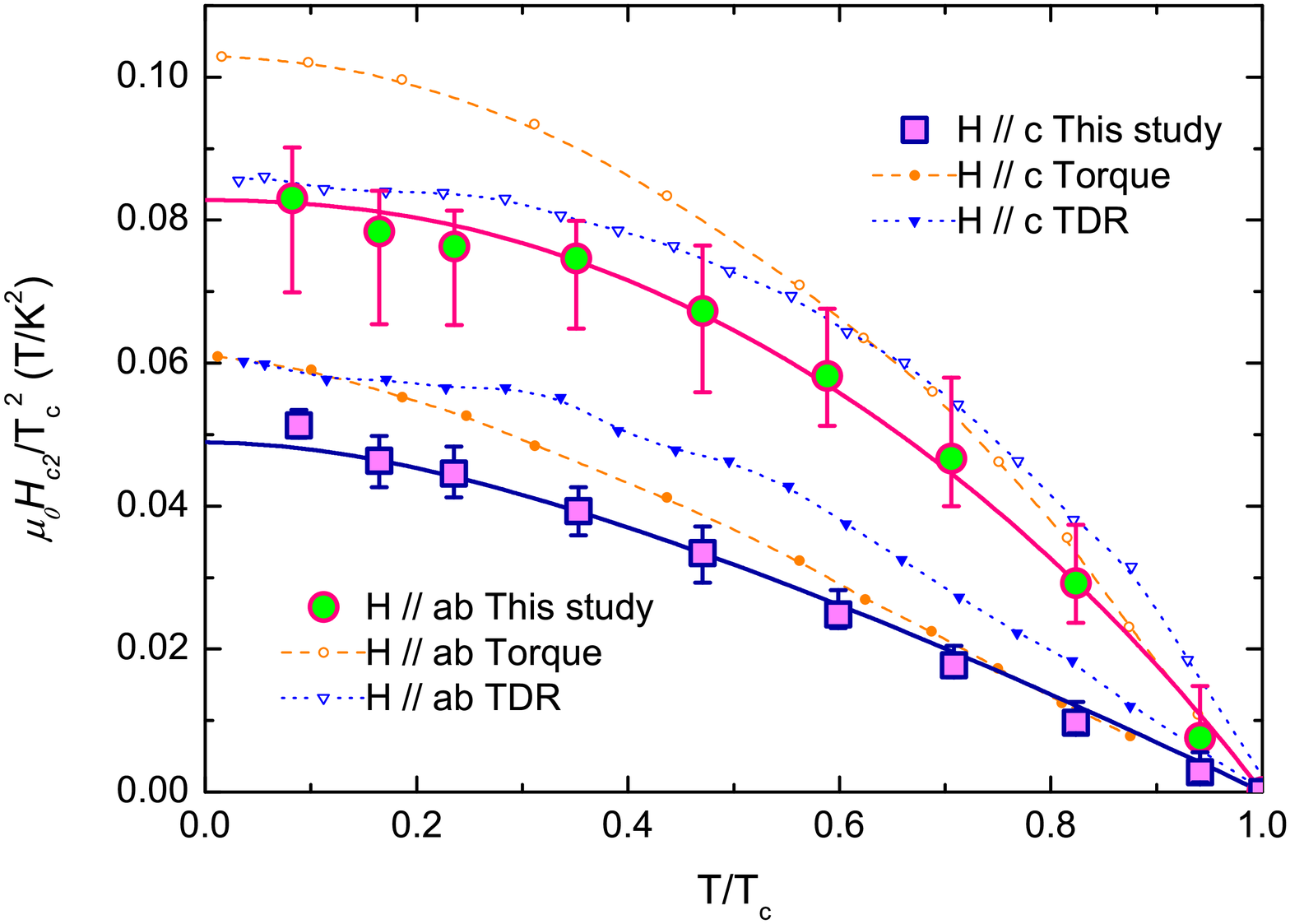}
\caption{The scaled upper critical field $\mu_0H_{c2}(T_c)/T_c^2$
versus the normalized temperature $T/T_c$ for LiFeAs. Symbols of the
square ($\square$), circle ($\circ$) and triangle ($\triangledown$)
represent the data obtained from measurements of the electrical
resistivity (this study), the magnetic torque \cite{KURITA} and the
resonant frequencies based on the tunnel-diode oscillator (TDR)
~\cite{Cho}, respectively.}\label{fig4}
 \end{figure}

\begin{table*}[tpb]
\centering \caption{\label{arttype}The derived superconducting
parameters for LiFeAs} \footnotesize\rm
\begin{tabular*}{\textwidth}{@{}l*{15}{@{\extracolsep{0pt plus12pt}}c}}

\hline
field&$T_c$(K)&-$\frac{d\mu_0H_{c2}}{dT}|_{T_c}$ (T/K)&$\mu_0H_{c2}$(1.4K)(T)&$\mu_0H_{c2}^{orb}$(T)&$\mu_0H_{c2}^P$(T)&$\alpha$&$\lambda_{so}$&$\xi$(nm)\\
\hline
$H\parallel c$&17.5&1.2&15&14.5&32.6&0&0&1.7\\
$H\parallel ab$&17.5&3.3&24.2&39.8&32.6&1.74&0.3&4.8\\
\hline

\end{tabular*}
\end{table*}

For conventional superconductors, $\mu_0H_{c2}^P(0)$ is usually much
larger than $\mu_0H_{c2}^{orb}(0)$ and, therefore, their upper
critical field is mainly restricted by the orbital pair-breaking
mechanism. In our case, the initial slope of $\mu_0H_{c2}$ at $T_c$,
i.e., -$d\mu_0H_{c2}/dT|_{T=T_c}$, is determined as 3.3 T/K and 1.2
T/K for $H\parallel ab$ and $H\parallel c$, respectively. Thus the
values of $\mu_0H_{c2}^{orb}(0)$ are accordingly derived as 39.8T
for $H\parallel ab$ and 14.5T for $H\parallel c$; the latter is
close to the experimental value of $\mu_0H_{c2}\simeq$ 15T at
$T=$1.4K, indicating an orbitally limited critical field for
$H\parallel c$. On the other hand, Eq. 2 yields
$\mu_0H^p_{c2}(0)=32.6$T. The experimentally derived value of
$\mu_0H_{c2}(0)\sim 25$T for $H\parallel ab$ is, therefore, well
below the corresponding values of $\mu_0H_{c2}^{orb}(0)$ and
$\mu_0H_{c2}^P(0)$. The solid lines in Fig. 3 present the WHH fits
to the experimental data of $\mu_0H_{c2}(T_c)$, in which both the
spin-paramagnetic and orbital pair-breaking effects were considered
\cite{WHH}. The parameter $\lambda_{so}$ describes the strength of
the spin-orbit scattering. The fits give the Maki parameter
$\alpha=$0 and 1.74 for field along the c-axis and the ab-plane,
respectively. The former further confirms the orbitally limited
critical field for $H\parallel c$. However, the resulted fitting
parameters ($\alpha=1.74, \lambda_{so}=0.3$) indicate that the upper
critical field is likely spin-paramagnetically limited for
$H\parallel ab$ even though we still could not exclude the
possibility of the orbital effect due to its multi-band effect. As
shown in Fig. 3 (see the dotted line and the solid line for
$H\parallel ab$), the spin-paramagnetic effect might lower the upper
critical field, and therefore, reduce the anisotropy of
$\mu_0H_{c2}$ at low temperatures.
For comparison, Fig. 4 plots the
available upper critical fields for LiFeAs, independently determined
from measurements of the electrical resistivity (this work), the
magnetic torque \cite{KURITA} and the resonant frequencies based on
a tunnel-diode oscillator \cite{Cho}. One can see that the
experimental results obtained from the above three methods are
similar in general; the visible discrepancy might result from the
exact determination of $T_c$. Nevertheless, the electrical
resistivity studied here provides the most direct approach for
determining the upper critical field.

\begin{figure}[htbp]\centering
 \includegraphics[width=8cm]{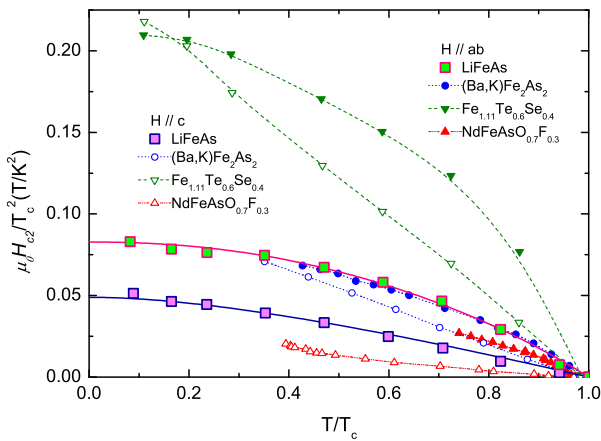}
\caption{The upper critical field $\mu_0H_{c2}/T_c^2$ versus the
normalized temperature $T/T_c$ for single crystals of LiFeAs (this
study), (Ba,K)Fe$_2$As$_2$ \cite{Yuan} and
Fe$_{1.11}$Te$_{0.6}$Se$_{0.4}$ \cite{Fang},
NdFeAsO$_{0.7}$F$_{0.3}$ \cite{Jaroszynski}, which $T_c$=17.5K, 55K,
28K and 14K, respectively. Note that variant symbols represent
variant compounds as marked in the figure. }\label{fig5}
 \end{figure}

In Fig. 5, we compare the upper critical field and its anisotropy in
several typical iron-based superconductors, i.e., LiFeAs (this
work), (Ba,K)Fe$_2$As$_2$ \cite{Yuan} and
Fe$_{1.11}$Te$_{0.6}$Se$_{0.4}$ \cite{Fang},
NdFeAsO$_{0.7}$F$_{0.3}$ \cite{Jaroszynski}. In general, the upper
critical fields of all these compounds show a rather weak anisotropy
at low temperatures in comparison with other layered
superconductors, e.g., the high $T_c$ cuprates and the organic
superconductors \cite{shriffer, singleton}. This indicates that the
inter-layer coupling might become significantly important in the
iron-based superconductors, which was ignored while modeling the
high $T_c$ cuprates. Among the iron-based superconductors, LiFeAs
shows a relatively small upper critical field. For example,
Fe$_{1.11}$Te$_{0.6}$Se$_{0.4}$ undergoes a superconducting
transition at $T_c\simeq 14$K, but it shows a much larger upper
critical fiel ($\mu_0H_{c2}(0)\simeq 45$T), which is likely
attributed to its higher disorder. In (Ba,K)Fe$_2$As$_2$ and
Fe$_{1.11}$Te$_{0.6}$Se$_{0.4}$ systems \cite{Yuan,Fang}, we
observed a nearly isotropic upper critical field at low temperature,
which unique feature was attributed to the three-dimensional-like
Fermi surface as experimentally confirmed later \cite{Vilmercati}.
The moderate anisotropy of $\mu_0H_{c2}$ in LiFeAs and the
1111-series is actually consistent with the band structure
calculations which indicate an enhanced anisotropy in these systems
\cite{Nekrasov}.

\section{Conclusion}

In summary, we have determined the complete temperature-magnetic
field phase diagram for the superconductor LiFeAs by means of
measuring the electrical resistivity in a field up to 40T. The upper
critical field of LiFeAs is derived as $\mu_0H_{c2}$(1.4K)=15T and
24.2T for field applied along the c-axis and the ab-plane,
respectively. The anisotropic parameter $\gamma$ decreases with
decreasing temperature and shows a weak anisotropic effect at low
temperatures. These findings indicate that weak anisotropy of
$\mu_0H_{c2}$ seems to be a common feature of the iron-based
superconductors, in spite of the layered nature of their crystal
structure.
\section{Acknowledgements}
This work was supported by the National Science Foundation of China
(No.10874146 and No. 10934005), the National Basic Research Program
of China (973 program) under grant No. 2011CBA00103 and
2009CB929104, the PCSIRT of the Ministry of Education of China,
Zhejiang Provincial Natural Science Foundation of China and the
Fundamental Research Funds for the Central Universities. Work at
NHMFL-LANL is performed under the auspices of the National Science
Foundation, Department of Energy and State of Florida.

\end{document}